\documentclass[aps,pra,twocolumn,groupedaddress,amsmath,amssymb]{revtex4}

\usepackage{epsfig,afterpage}
\usepackage{graphicx}% Include figure files
\usepackage{amsfonts}
\usepackage{amssymb}
\usepackage{indentfirst}
\usepackage{amsmath,amsthm}
\usepackage{dsfont}

\usepackage{epsfig}

\usepackage{multirow}
\usepackage{pstricks}
\usepackage{lscape}
\def\beq{\begin{equation}}
\def\eeq{\end{equation}}
\def\bea{\begin{eqnarray}}
\def\eea{\end{eqnarray}}
\def\bq{\begin{quote}}
\def\eq{\end{quote}}
\def\ben{\begin{enumerate}}
\def\een{\end{enumerate}}
\def\bit{\begin{itemize}}
\def\eit{\end{itemize}}
\def\nn{\nonumber}

\newcommand{\product}[1]{\ensuremath{{\cal P}{#1}}}
\newcommand{\productpar}[1]{\ensuremath{{\cal P}{#1}_{\pi}}}
\newcommand{\separ}[1]{\ensuremath{{\cal S}{#1}}}
\newcommand{\separpar}[1]{\ensuremath{{\cal S}{#1}_{\pi}}}
\newcommand{\proj}[2]{\ensuremath{{\mathbb P}_{#1}^{#2}}}

\newcommand{\equivsep}[1]{\ensuremath{[{\cal S}{#1}_{\pi}}]}

\def\parity{\hat{P}} % parity operator
\def\defeq{:=}
\def\pset{\Pi}  % set of conserving parity states
\def\antipi{\not\pi}

\newcommand{\PinPiisym}{\ensuremath{\rho_{\product{1}}}}

\def\PinPii{\ensuremath{\frac{1}{16}\left(
\begin{array}{rrrr}
9 & 0 & 0 & -i\\
0 & 3 & -i & 0 \\
0 & i & 3 & 0\\
i & 0 & 0 & 1
\end{array}
\right)}}

\newcommand{\PiinPiiisym}{\ensuremath{\rho_{\product{2}}}}

\def\PiinPiii{\ensuremath{\frac{1}{4}\left(
\begin{array}{rrrr}
1 & 1 & -1 & -1\\
1 & 1 & -1 & -1\\
-1 & -1 & 1 & 1\\
-1 & -1 & 1 & 1
\end{array}
\right)}}

\newcommand{\PiiinPiisym}{\ensuremath{\rho_{\product{3}}}}

\def\PiiinPii{\ensuremath{\frac{1}{6} \left (
\begin{array}{rrrr}
2 & 2 & 0 & 0\\
2 & 2 & 0 & 0\\
0 & 0 & 1 & -1\\
0 & 0 & -1 & 1
\end{array}
\right )}}

\newcommand{\ZinSisym}{\ensuremath{\rho_{\equivsep{1}}}}

\newcommand{\ZinSi}{\ensuremath{\frac{1}{15}\left(
\begin{array}{cccc}
5 & 0 & 0 & 2\sqrt{5}\\
0 & 3 & 3 & 0\\
0 & 3 & 3 & 0\\
2\sqrt{5} & 0 & 0 &4\end{array}\right)}}

\newcommand{\SinSiisym}{\ensuremath{\rho_{\separpar{1}}}}

\newcommand{\SiipnSiisym}{\ensuremath{\rho_{\separpar{2'}}}}

\newcommand{\SiipnSii}{\ensuremath{\frac{1}{4}\left(
\begin{array}{cccc}
1 & 0 & 0 & 0\\
0 & 1 & 1 & 0\\
0 & 1 & 1 & 0\\
0 & 0 & 0 & 1 \end{array}\right)}}

\begin{document}

\title{Entanglement in fermionic systems}
\author{Mari-Carmen Ba\~nuls, J. Ignacio Cirac, Michael M. Wolf}
\affiliation{Max-Planck-Institut f\"ur Quantenoptik,
Hans-Kopfermann-Str. 1, 85748 Garching, Germany.}
\date{\today}

\begin{abstract}

The anticommuting properties of fermionic operators, together with
the presence of parity conservation, affect the concept of
entanglement in a composite fermionic system. Hence different
points of view can give rise to different reasonable definitions
of separable and entangled states. Here we analyze these
possibilities and the relationship between the different classes
of separable states. We illustrate the differences by providing a
complete characterization of all the sets defined for systems of
two fermionic modes. The results are applied to Gibbs states of
infinite chains of fermions whose interaction corresponds to a
XY-Hamiltonian with transverse magnetic field.

\end{abstract}

\maketitle

% Introduction (motivation?) should be maybe in a previous paragraph
\section{Introduction}

The definition of entanglement in a composite quantum system
~\cite{Werner1} depends on a notion of locality which is typically
assigned to a tensor product structure or to commuting sets of
observables~\cite{Werner2}. 
Various a priori different definitions
can then be formulated depending on requirements concerning
preparation, representation, observation and application.
Fortunately, most of them usually coincide.
\cite{foot1}
%\cite{Yet a different definition is based on a generalized notion 
%of entanglement which is relative to distinguished subspaces 
%of observables~\cite{barnum04}.}

In the present article we investigate systems of fermions where
several of these definitions differ due to indistinguishability,
anti-commutation relations and the parity superselection rule. We
will provide a systematic study of the different definitions of
entanglement and determine the relations between them. To this
end, we will consider fermionic systems in second quantization.
That is, the entanglement will be between sets of modes or regions
in space rather than between particles. The latter case was
studied in first quantization in~\cite{schliemann01,li01,levay05}
whereas entanglement between distinguishable modes of fermions has
been calculated for various systems in
~\cite{zanardi02,zanardi02iti,larsson06,wolf06area,GioevKlich,cramer06}.

That the presence of superselection rules affects the concept of
entanglement has been pointed out and studied in detail in
~\cite{verstraete03,schuch04ssr,schuch04ssr2,wiseman03b,caban05}.
There, the existence of states was shown which are convex
combinations of product states but not locally preparable, thus
two reasonable definitions of entanglement already differ.

In the following the differences will mainly arise from an
interplay between the parity superselection rule and the
anti-commutation relation of fermionic operators. The different
mathematical definitions will carry physically motivated meanings
 corresponding to different
abilities to prepare, use or observe the entanglement, as well as
to differences between the single copy case and the asymptotic
regime.

%In the rest of the paper ...
In section~\ref{preliminaries} we introduce the basic ideas and
tools used in the rest of the paper. We start by defining the
different sets of product states in section~\ref{products}. From
them, several sets of separable states are constructed by convex
combination in section~\ref{separable}. It is shown that they all
correspond to four different classes each of which contains the
previous ones as proper subsets:
\begin{enumerate}
    \item States which are preparable by means of local operations
    and classical communication (LOCC).
    \item Convex combinations of product states in Fock space.
    \item Convex combinations of states for which products of
    locally measurable observables factorize.
    \item States for which all locally measurable correlations can
    as well arise from a state within class 3 above.
\end{enumerate}

 Section~\ref{multiple}
analyzes the asymptotic properties of the various sets of
separable states. As an illustration of all these concepts,
section~\ref{sec:1x1} shows the complete characterization of the
different sets in the case of a $1\times1$-modes system, and their
application to the thermal state of an infinite chain of fermions
interacting with a particular Hamiltonian. 
In order to improve the readability of the paper,
we have compiled the detailed proofs of all the relations
in section~\ref{sec:proofs}.
%Finally, the detailed
%proofs of all the relations among the different sets are derived
%in section~\ref{sec:proofs}.

\section{Preliminaries}
\label{preliminaries}
% define symbols, Gaussian states, parity operators, fermionic states

The basic objects for describing a fermionic system
of $m$ modes are the creation and annihilation operators,
which satisfy canonical anticommutation relations.
Alternatively, $2 m$ Majorana operators can be
defined, $c_{2k-1} \defeq a_k^\dagger+a_k$,
$c_{2k} \defeq (-i)(a_k^\dagger-a_k)$, for $k=1,\ldots 2m$,
which satisfy $\{c_i,c_j\}=\delta_{ij}$.
Each set generates the algebra ${\cal C}$ of all observables.
A bipartition of the system is defined by two subset of modes,
$A=1,\ldots\,m_A$ and $B=m_A+1,\ldots\,m$.
We will denote by $\cal A$ ($\cal B$) the operator subalgebra
spanned by the $m_A$ ($m_B$) modes in $A$ ($B$).

If $n_k$ is the occupation number of the $k$-th mode, i.e. the
expectation value of the operator
$a_k^{\dagger} a_k$,
%%=\frac{1}{2}\left(\mathds{1}+\frac{i}{2}[c_{2 k},\,c_{2k-1}] \right),$$
the Fock basis can be defined by \beq \lvert n_1,\ldots\,n_m
\rangle
=(a_1^{\dagger})^{n_1}\ldots(a_m^{\dagger})^{n_m}\lvert0\rangle.
\eeq The Jordan-Wigner transformation maps the fermionic algebra
onto Pauli spin operators so that \beq
c_{2k-1}=\prod_{i=1}^{k-1}\sigma_z^{(i)} \sigma_x^{(k)},\quad
c_{2k}=\prod_{i=1}^{k-1}\sigma_z^{(i)} \sigma_y^{(k)}.
\label{eq:JW} \eeq The Hilbert space associated to $m$ fermionic
modes (Fock space) is isomorphic to the $m$-qubit space. Due to
the anticommutation relations, however, the action of fermionic
operators in Fock space is non-local.

For the fermionic systems under consideration, conservation of
the parity of the fermion number, $\parity = i^m \prod_k c_k$,
implies that the accessible state space is the
direct sum of positive (even) and negative (odd) parity
eigenspaces.
Any physical state or observable commutes with the operator $\parity$,
so that we can define the set of physical states
$$\pset \defeq \{\rho : [\rho,\,\parity]=0 \}.$$
Correspondingly,
$\cal A_\pi$ and $\cal B_\pi$ will designate
the sets of local observables, commuting with the local parity operators
$\parity_A$ and $\parity_B$, respectively.
%In the same way, $\cal A_\antipi$ and $\cal B_\antipi$ will denote the rest of the operator subalgebras.

We will call an observable even if it commutes with the parity operator,
whereas an odd observable will be that anticommuting with $\parity$.
Notice that with this nomenclature, odd observables are not the ones
supported on the odd parity eigenspace.
On the contrary, an observable with such support
will be even in this notation, as it commutes with $\parity$.
It will be convenient to make use of the projectors onto the well-defined
parity subspaces, \proj{e(o)}{}. Any state (or operator) commuting with parity
has a block diagonal structure
$\rho=\proj{e}{}\rho\proj{e}{}+\proj{o}{}\rho\proj{o}{}$.
In the local subspaces, a parity conserving operator can be written
$A_\pi=\proj{e}{A}A_\pi\proj{e}{A}+\proj{o}{A}A_\pi\proj{o}{A}$.

One subset of states of particular physical interest is that of
Gaussian states. They describe the equilibrium and excited states
of quadratic Hamiltonians. Moreover, important variational states
(e.g. the BCS state) belong to this category. In various respects
Gaussian states exhibit relevant extremality
properties~\cite{wolf06extremality,wolf07extremality}. Fermionic
Gaussian states are those whose density matrix can be written as
an exponential of a quadratic form in the fermionic
operators~\cite{bravyi05},
$$
\rho=\exp \left(-\frac{i}{4} c^T M c\right),
$$
for some real antisymmetric matrix $M$.
The covariance matrix of any fermionic state is a real antisymmetric
matrix defined by
$$
\Gamma_{k l}=\frac{i}{2} \mathrm{tr} \left(\rho[c_k,\,c_l]\right),
$$
which necessarily satisfies $i \Gamma \leq \mathds{1}$. According
to Wick's theorem, the covariance matrix determines completely all
the correlation functions of a Gaussian state. Pure fermionic
Gaussian states satisfy $\Gamma^2=-\mathds{1}$, and they can be
written as a tensor product of pure states involving at most one
mode of each partition~\cite{botero04bcs}.

\section{Product states}
\label{products}
% define three sets of product states, intersect with parity
% show inclusions

%% Detailed definitions of product states

We start by defining product states of a bipartite fermionic system
formed by $m=m_A+m_B$ modes, where $m_A$ ($m_B$) is the number of modes
in partition $A$ ($B$).

The entanglement of such system can be studied at the level of operator
subalgebras
or in the Fock space representation, thus the possibility to define
different sets of product states.
In Fock space, the isomorphism to a system of
$m_A+m_B$ qubits allows separability to be studied with respect to
the tensor product $\mathbb{C}^{2m_A}\otimes\mathbb{C}^{2m_B}$.
At the level of the operator subalgebras, on the other hand,
one should study the entanglement between ${\cal A}$ and ${\cal B}$
subalgebras.
% REWRITE!!! --->
However the observables in them do not
commute, in general, and have non-local action in Fock
space. % <---- REWRITE!!!
On the contrary, ${\cal A}_\pi$ and ${\cal B}_\pi$, i.e. the subalgebras of
parity conserving operators, commute with each other,
so that they can be considered local to both parties.
It is then natural to study the entanglement between them.

\subsection{General states}

With these considerations, we may give the following definitions
of a product state. They are summarized in Table~\ref{tab:prod}.
%%%%%%%%%%%%%%%%%%%%%%%%%%%%%%%%%%%%%%%%%%%%%%%%%%%%%%%%%%%%%%%%
% Table: Product states
\begin{table*}
\begin{tabular}{|c|c|c|c|}
\hline
Set& Definition & Relation & Example in $1\times1$ \\
\hline
\multirow{2}{*}{
\product{0}
}
&
\multirow{2}{*}{
$\rho(A_\pi B_\pi)=\tilde{\rho}(A_\pi B_\pi),\, \tilde{\rho}\in\product{2}$
}
&
% aqui estaba P0=P1
&
\multirow{4}{*}[.5ex]{
$\PinPiisym=\PinPii \in \productpar{1}\setminus \productpar{2}
$
}
\\
&&
\multirow{2}{*}{
$\product{0}=\product{1}$
}

&
\\
\cline{1-2}
\multirow{2}{*}{
\product{1}
}
&
\multirow{2}{*}{
$\rho(A_\pi B_\pi)=\rho(A_\pi)\rho(B_\pi)$
}
&
&

\\
\cline{3-3}
&&
\multirow{2}{*}{
$\product{2}\subset\product{1}$
}
&
\\
\cline{1-2}\cline{4-4}
\multirow{4}{*}{
\product{2}
}
&
\multirow{4}{*}{
$\rho=\rho_A\otimes\rho_B$
}
&
&
\multirow{4}{*}[.5ex]{
$\PiinPiiisym=\PiinPiii \in \product{2}\setminus \product{3}
$
}\\
\cline{3-3}
&&&\\
&&&
\\
&&
\multirow{2}{*}{
$\product{2}\neq\product{3}$
}
&
\\
\cline{1-2}
\cline{4-4}
\product{3}&
$\rho(A\,B)=\rho(A)\rho(B)$&
&
$
\PiiinPiisym=\PiiinPii \in \product{3}\setminus \product{2}
$
\\
\hline
\end{tabular}
\caption{Relations among sets of product states.}
\label{tab:prod}
\end{table*}

\bit
\item
We may call a state product if there exists some state
acting on the Fock space
of the form
$\tilde{\rho}=\tilde{\rho}_A\otimes\tilde{\rho}_B$,
and producing the same expectation values for all local observables.
Formally,~\cite{foot2}
%\footnote{In the following we use for the expectation values the notation 
%$\rho(X)\defeq\mathrm{tr}(\rho\,X)$.}
\bea
\nn
\product{0} &:=& \left\{ \rho :\,  \exists\tilde{\rho}_A,\,\tilde{\rho}_B, \, [\tilde{\rho}_{A(B)},\parity_{A(B)}]=0\quad \mathrm{s.t.} \right.\\
&&\left. \quad \rho(A_\pi \, B_\pi) =\tilde{\rho}_A(A_\pi)\tilde{\rho}_B(B_\pi) \right.\\
\nn
&&\left. \quad \quad\forall  A_\pi\in {\cal A}_{\pi},\, B_\pi\in {\cal B}_{\pi} \right\}.
\eea

\item
Alternatively, product states may be defined as those for which
the expectation value of products of local observables factorizes,
\bea
\nn
\product{1} &:=& \left\{ \rho : \rho(A_\pi \, B_\pi) =\rho(A_\pi) \rho(B_\pi)
\right.\\
&&\left. \quad \quad  \forall A_\pi\in {\cal A}_{\pi},\, B_\pi\in {\cal B}_{\pi} \right\}.
\eea

\item
At the level of the Fock representation, a product state
can be defined as that writable as a tensor product,
\beq
\product{2} :=  \left\{ \rho : \rho=\rho^A \otimes \rho^B \right\}.
\eeq

\item
From the point of view of the subalgebras of observables for both
partitions, one may ignore the commutation with the parity operator
and require factorization of any product of observables for a product
state~\cite{moriya05}. This yields another set
\beq
\product{3} :=\left\{ \rho : \rho(A \, B) =\rho(A) \rho(B)
\quad  \forall A\in {\cal A},\, B\in {\cal B} \right\}.
\eeq

\eit

% Relation between them?

The two first definitions are equivalent,
$\product{0}\equiv\product{1}$.
% REWRITE --->
They correspond to states
with a separable projection onto the diagonal blocks that preserve
parity in each of the subsystems. % <--- REWRITE
This means that
$$\sum_{\alpha,\,\beta=e,\,o} \proj{\alpha}{A}\otimes \proj{\beta}{B}
\rho
 \proj{\alpha}{A}\otimes \proj{\beta}{B},
$$
is a product in the sense of \product{2}.

The three remaining sets are strictly different.
In particular
$\product{2} \subset \product{1}$
and
$\product{3} \subset \product{1}$,
but $\product{3} \neq \product{2}$.
%% Alternative: $\product{2}\cup \product{3} \subset \product{1}$.
The inclusion $\product{2},\product{3} \subseteq \product{1}$
is immediate from the definitions.
The non equality of the sets can be seen by explicit examples as those
shown in Table~\ref{tab:prod}.
The difference between $\product{3}$ and $\product{2}$, however,
is limited to non-physical states, i.e. those
not commuting with parity~\cite{moriya05}.

\subsection{Physical states}
% Parity commuting states

Being parity a conserved quantity in the systems of interest,
the only physical states will be those commuting with $\parity$.
It makes then sense to restrict the study of entanglement to such
states.
By applying each of the above definitions to the physical states,
$\pset$, we obtain the following sets of physical product states.

\bit
\item
$\productpar{1} := \product{1} \cap \pset= \product{0} \cap \pset$
\item
$\productpar{2} := \product{2} \cap \pset$
\item
$\productpar{3} := \product{3} \cap \pset$
\eit
We notice that
$\rho\in\productpar{2}$ is equivalent to $\rho=\rho_A\otimes\rho_B$
where both factors are also
parity conserving.

With the parity restriction, the three sets are related by
\beq
\productpar{3}=\productpar{2}\subset\productpar{1}.
\eeq

The proofs of all the relations above 
are shown in section~\ref{subsec:prod}.

%%%%%%%%%% Proofs %%%%%%%%%%%%%

%The relation $\productpar{3} \subseteq \productpar{1}$ is immediate, as
%any $B \in {\cal B}_\pi$ is mapped by the Jordan-Wigner transformation
%to an even operator in $\mathds{C}^{2 m_B}$.
%The strict inclusion can be seen with an explicit example.
%% % Contar el ejemplo!!
%% Let us consider, for a 2-mode system, the Gaussian state
%% defined by the covariance matrix
%% \beq
%% \Gamma=\left (
%% \begin{array}{c c c c}
%% 0 & a & x & 0 \\
%% -a & 0 & 0 & 0\\
%% -x & 0 & 0 & b\\
%% 0& 0 & -b & 0
%% \end{array}
%% \right ), \quad
%% \begin{array}{l}
%% 0 < |x|\leq 1, \\
%% |a| \leq \sqrt{1-x^2},\\
%% |b|\leq \sqrt{1-\frac{x^2}{1-a^2}}.
%% \end{array}
%% \label{exp2inp1}
%% \eeq
%% For such a small system, the definition of $\product{1}$
%% reduces to a single relation between expectation values,
%% written in terms of Majorana operators as $\langle c_1  c_2 c_3 c_4\rangle = \langle  c_1 c_2\rangle \langle c_3 c_4 \rangle$.
%% The state~\ref{exp2inp1} satisfies the condition and thus
%% belongs to $\productpar{1}$.
%% However, it is not in $\productpar{2}$, as it yields a non-vanishing
%% expectation value for $\langle c_1 c_3 \rangle$.
%% Vanishing of any expectation value of the form
%% $\langle A_{\antipi}B_{\antipi}\rangle$ is a necessary condition
%% for  $\rho \in \productpar{2}$ (\cite{moriya05}).
%% As a consequence, the covariance matrix of any such state
%% will be block-diagonal, $\Gamma=\Gamma_A\oplus\Gamma_B$.

%%%%%%%% End of Proofs %%%%%%%%%%%5

\subsection{Pure states}
% Pure states!!!

For pure states, all $\productpar{i}$ reduce to the same set.
If the state vector is written in a basis of well-defined parity
in each subsystem,
%$\left\{\lvert e_i \varepsilon_j\rangle,\lvert e_i \theta_j\rangle,\lvert o_i \varepsilon_j\rangle,\lvert o_i \theta_j\rangle \right\}$,
it is possible to show that the condition of
$\productpar{1}$ requires that such expansion has a single
non-vanishing coefficient, and thus the state can be written as
a tensor product also with the definition of $\product{2}$.
%(see section~\ref{sec:proofs} for a complete proof).

\section{Separable states}
\label{separable}
% take convex hulls of the above sets (before and after intersecting
% with parity
% Define equivalence classes as equality of local parity conserving operators
% (locally undistinguishable)=> new set

\begin{figure}
\resizebox{\columnwidth}{!}{
\input{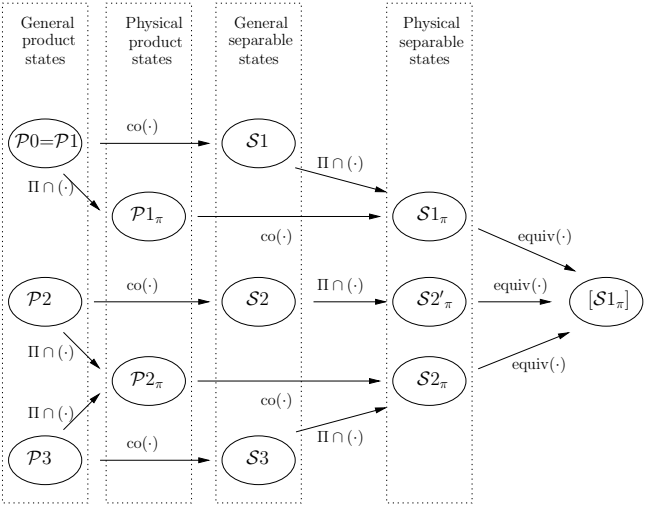}
}
\caption{Scheme of the construction of the different sets.}
\label{fig:scheme}
\end{figure}

Generally speaking, separable states are those that can be written
as convex combination of product states. The convex hulls of the
different sets of product states introduced in the previous section
define then various separability sets. Fig.~\ref{fig:scheme}
outlines the procedure to obtain each of these sets. Table
~\ref{tab:separ} summarizes the different definitions and the
relations between them.

\subsection{General states}

Taking the convex hull of the general product states,
we define the sets
\bit
\item
$\separ{1}\defeq\mathrm{co}\left(\product{1}\right)$,
\item
$\separ{2}\defeq\mathrm{co}\left(\product{2}\right)$,
\item
$\separ{3}\defeq\mathrm{co}\left(\product{3}\right)$.
\eit

These contain both physical states, commuting with $\parity$,
and non-physical ones.
It can be shown that
$\separ{3} \subset \separ{2} \subset \separ{1}$.

% Proofs at the end.
The non-strict inclusion $\separ{2} \subseteq \separ{1}$
is immediate from the inclusion between product sets.
The strict character can be seen with an example, in particular
in the subset of physical states, $\SinSiisym$.
$\separ{3}\subset \separ{2}$ was proved in~\cite{moriya05}.

\subsection{Physical states}

From the physical sets of product states we define the
following sets of separable states,
\bit
\item
$\separpar{1}\defeq\mathrm{co}\left(\productpar{1}\right)$,
\item
$\separpar{2}\defeq\mathrm{co}\left(\productpar{2}\right)$,
\eit
Obviously, the corresponding $\separpar{3}\equiv\separpar{2}$.
The inclusion relations among product states imply
$\separpar{2}\subseteq\separpar{1}$.
It is easy to see with an example that this inclusion is also strict.
Table~\ref{tab:separ} summarizes the definitions and mutual relations
of the various separability sets.

%%%%%%%%%%%%%%%%%%%%%%%%%%%%%%%%%%%%%%%%%%%%%%%%%%%%%%%%%%%%%%%%
% Table: Separable states

% useful length
\newlength{\examplewidth}
\settowidth{\examplewidth}{$\ZinSi \in \equivsep{1}\setminus\separpar{1}$}

\begin{table*}
\begin{tabular}{|c|c|c|c|c|}
\hline
Set& Definition & Characterization & Relations & Example in $1\times1$ \\
\hline
%%%%%%%%%%%%%%%%%%%%%%%%% [S1pi]
\multirow{4}{*}{
\equivsep{1}
}&
% Def?
\multirow{4}{*}{
\equivsep{1}
}
&
\multirow{4}{*}{
$
\sum_{\alpha,\,\beta=e,\,o} \proj{\alpha}{A}\otimes \proj{\beta}{B}
\rho
 \proj{\alpha}{A}\otimes \proj{\beta}{B} \in \separpar{2'}
$
}
&
&
%example
\multirow{4}{*}[.5ex]{
$
\ZinSisym = \ZinSi
\in\equivsep{1}\setminus\separpar{1}
$
}
\\
%second line of [S1]
&&&&
\\
%third line of [S1]
&&&&
\\
%fourth line of [S1]
&&&
\multirow{2}{*}{
$\separpar{1}\subset\equivsep{1}$
}
&\\
\cline{1-3}\cline{5-5}
%%%%%%%%%%%%%%%%%%%%%%%%% S0
\multirow{2}{*}{
\separpar{0}
}
&
\multirow{2}{*}{
$\mathrm{co}(\productpar{0})$
}
&
\multirow{4}{*}{
$
\begin{array}{c}
\rho=\sum_k \lambda_k \rho_k, \quad
 s.t. \\
\sum_{\alpha,\,\beta=e,\,o} \proj{\alpha}{A}\otimes \proj{\beta}{B}
\rho_k
 \proj{\alpha}{A}\otimes \proj{\beta}{B} \in \separpar{2'}
\end{array}
$
}
&
&
%example
\multirow{4}{\examplewidth}{
For the $1\times 1$ case, $\separpar{1}=\separpar{2'}$. Therefore examples of $\separpar{1}\setminus\separpar{2'}$ can only be found in bigger systems, f.i. $2\times 2$ modes.
}
\\
\cline{4-4}
% Second row of S0
&&&
\multirow{2}{*}{
$\separpar{1}=\separpar{0}$
}
&\\
\cline{1-2} %\cline{5-5}
%%%%%%%%%%%%%%%%%%%%%%%%% S1
\multirow{2}{*}{
\separpar{1}
}&
\multirow{2}{*}{
$\mathrm{co}(\productpar{1})$
}
&

&
%example
&
\\
\cline{4-4}
% Second row of S1
&&&
\multirow{2}{*}{
$\separpar{2'}\subset\separpar{1}$
}
&
\\
\cline{1-3}\cline{5-5}
%%%%%%%%%%%%%%%%%%%%%%%%% S2'
\multirow{4}{*}{
\separpar{2'}
}&
\multirow{4}{*}{
$\mathrm{co}(\product{2})\cap\pset$
}
&
\multirow{4}{*}{
$\rho=\sum_{k} \lambda_k \rho_k^A\otimes\rho_k^B$
}&
&
\multirow{2}{*}[.5ex]{
$\SiipnSiisym=\SiipnSii \in \separpar{2'} \setminus \separpar{2}$
}
\\
\cline{4-4}
% Second row S2'
&&&&\\
%Third row S2'
&&&&\\
% Last row S2'
&&&
\multirow{2}{*}{
$\separpar{2}\subset\separpar{2'}$
}
&
\\
\cline{1-3}\cline{5-5}
%%%%%%%%%%%%%%%%%%%%%%%%% S2
\multirow{2}{*}{
\separpar{2}
}&
\multirow{2}{*}{
$\mathrm{co}(\productpar{2})$
}
&
% character
\multirow{2}{*}{
$\proj{e}{}\rho\proj{e}{}\in\separpar{2'}$, $\proj{o}{}\rho\proj{o}{}\in\separpar{2'}$
}
&&
\multirow{2}{*}{
$\rho \in \separpar{2} \Leftrightarrow \rho$ diagonal
}
\\
% Second row S2
&&&&
\\
\hline
\end{tabular}
\caption{The different sets of separable states and their
relations. $\separpar{2}$ contains all states preparable by means
of LOCC. $\separpar{2'}$ represents the original definition of
separability in Fock space. $\separpar{1}$ gives all convex
combinations of states for which expectation values of products of
locally measurable observables factorize. $\equivsep{1}$ contains
all states which are locally indistinguishable from
$\separpar{1}$.
 } \label{tab:separ}
\end{table*}

As shown in Fig.~\ref{fig:scheme}, we may take the physical states
that satisfy the definitions for separability introduced in
the previous subsection, and hence use
$\separ{i}\cap\pset$ as the definition of separable states.
This yields the sets
\bit
\item
$\separ{1}\cap\pset\equiv\separpar{1}$,
\item
$\separpar{2'}\defeq\separ{2}\cap\pset$,
\item
$\separ{2}\cap\pset\equiv\separpar{2}$.
\eit

% Again, proofs somewhere else?
Only $\separpar{2'}$ is different from the
separable sets defined above.
Actually, given an $\separ{1}$ state that commutes with $\parity$,
it is possible to construct a decomposition according to $\separpar{1}$
by taking the parity preserving part of each term in the original
convex combination. 
Therefore $\separ{1}\cap\pset\subseteq\separpar{1}$, while the
converse inclusion is evident.
For $\separ{3}\cap\pset$, on the other hand, it was shown in~\cite{moriya05}
that any parity preserving state in $\separ{3}$ has a decomposition in
terms of only parity preserving terms, and is thus in $\separpar{3}$.

All the considerations above leave us with three strictly different
sets of separable physical states,
\beq
\separpar{2} \subset \separpar{2'} \subset \separpar{1}
\eeq

%%%%%%%%%%%%%%%%% AQUI AQUI
From the definitions, it
is immediate that $\separpar{2} \subseteq \separpar{2'}$.
The inclusion is strict because
not every state $\rho \in\separpar{2'}$
has a decomposition in terms of products of even states
(see example $\SiipnSiisym$ in Table~\ref{tab:separ}).
The condition for  $\separpar{2}$ is then more restrictive.

From the relation between product sets,
$\separ{2} \subseteq \separ{1}$,
and $\separpar{2'} \subseteq \separpar{1}$.
The strict inclusion can be shown by constructing an explicit example
of a $\productpar{1}$ state without positive
partial transpose (PPT)~\cite{peres96}
in the 2$\times$2-modes system.

The detailed proofs of 
the equivalences and inclusions above
are shown in section~\ref{subsec:separ}.

\subsection{Equivalence classes}

If one is only interested in the measurable correlations of the
state, rather than in its properties after further evolution or
processing, it makes sense to define an equivalence relation
between states by
$$
\rho_1 \sim \rho_2 \quad \mathrm{if} \quad  
\rho_1(A_\pi  B_\pi)=\rho_2(A_\pi B_\pi) \ \ \forall A_\pi\!\in\! \cal{A}_\pi,\,
B_\pi\!\in\!\cal{B}_\pi,
$$
i.e. two states are equivalent if they produce the same
expectation values for all physical local operators. Therefore,
two states that are equivalent cannot be distinguished by means of
local measurements.

With the restriction of parity conservation, the states that can be
locally prepared are of the form $\separpar{2}$, i.e.
$\rho=\sum_k \rho_k^A \otimes \rho_k^B$, where
$[\rho_k^{A(B)},\parity_{A(B)}]=0$.
Since the only
locally accessible observables are local, parity preserving
operators, i.e. quantities of the form $\rho(A_\pi B_\pi)$,
it makes sense to say that a given
state is separable if it is equivalent to a  state that can
be prepared locally.
With this definition, the set of separable states is equal to
the equivalence class of $\separpar{2}$ with respect to the
equivalence relation above.

Generalizing this concept, we may construct the equivalence classes
for each of the relevant separability sets,
$$
\equivsep{i}\defeq \{\rho\,:\, \exists \tilde{\rho}\in\separpar{i},\,
\rho\sim\tilde{\rho}\},\quad i=1,2',2.
$$
From the inclusion relation among the separability sets,
$\equivsep{2}\subseteq \equivsep{2'}\subseteq\equivsep{1}$.
And, obviously, $\separpar{i}\subseteq\equivsep{i}$.

On the other hand, any state $\rho\in\equivsep{1}$ has also an equivalent state
in $\separpar{2}$ (see section~\ref{subsec:separ}), so that
$$
\equivsep{2}=\equivsep{2'}=\equivsep{1}.
$$

This equivalence class includes then all the separability sets
described in the previous subsection.
However, it is strictly larger, as can be seen by the explicit example
$\ZinSisym$ in Table~\ref{tab:separ}.

\subsection{Characterization}

It is possible to give a characterization of the previously defined
separability sets in terms of the usual mathematical concept of
separability, i.e. with respect to the tensor product.
This allows us to use standard separability criteria 
(see~\cite{horodecki07review} for a recent review) in order to decide
whether a given state is in each of these sets.

The definition $\separpar{2'}$ corresponds
to the separability
in the sense of the tensor product, i.e. the standard notion~\cite{Werner1},
applied to parity preserving states.

As convex hull of $\product{2}\cap\pset$, the
set $\separpar{2}$ consists of states with a decomposition
in terms of tensor products, with the additional restriction that
every factor commutes with the local version of the parity operator.
Using the block diagonal structure
$\proj{e}{} \rho \proj{e}{} +\proj{o}{} \rho \proj{o}{}$
of any parity preserving state,
each block
must have independent decompositions in the sense of the tensor product.
Then a state will be in $\separpar{2}$ iff both
$\proj{e}{} \rho \proj{e}{}$ and $\proj{o}{} \rho \proj{o}{}$
are in $\separpar{2'}$.

A state $\rho$ is in $\productpar{0}$ if its diagonal blocks
are a tensor product,
\beq
\sum_{\alpha,\,\beta=e,\,o} \proj{\alpha}{A}\otimes \proj{\beta}{B}
\rho
 \proj{\alpha}{A}\otimes \proj{\beta}{B}=
\tilde{\rho}^A \otimes \tilde{\rho}^B \in \productpar{2}.
\label{charactP0}
\eeq
The set $\separpar{1}$ is characterized as
the convex hull of $\productpar{1}\equiv\productpar{0}$, i.e.
it is formed by convex combinations of states that
can be written as the sum of a parity preserving tensor product
plus some off--diagonal terms.

Finally, the equivalence class $\equivsep{1}\equiv\equivsep{2}$
is completely defined in terms of the expectation values of observable
products $A_\pi B_\pi$.
These have no contribution from off-diagonal blocks in $\rho$,
so the class can be characterized in terms of
the diagonal blocks alone.
Therefore a state is in $\equivsep{1}$ iff
\beq
\sum_{\alpha,\,\beta=e,\,o} \proj{\alpha}{A}\otimes \proj{\beta}{B}
\rho
 \proj{\alpha}{A}\otimes \proj{\beta}{B}
\in \separpar{2'}.
\label{charactZ1}
\eeq
Since the condition involves only the block diagonal part of the state,
it is equivalent to the individual separability
(with respect to the tensor product) of each of the blocks.

\section{Multiple copies}
\label{multiple}
% Asymptotic properties...

% Stable criteria, 
% OPEN: S1tilde???

% Que pasa con varias copias 
%% Motivacion??

The definitions introduced in the previous sections apply to 
a single copy of the fermionic state.
It is nevertheless interesting to 
see the stability of the different criteria when several copies 
are considered, and, in particular, to understand their asymptotic 
behaviour when $N\rightarrow\infty$.

% S2',S2 igual, incluso con 2
The criteria $\separpar{2'}$ and $\separpar{2}$ are stable
when several copies of the state are considered.
\bit
\item
$\rho^{\otimes 2}\in\separpar{2'} \iff \rho\in\separpar{2'}$,
\item
$\rho^{\otimes 2}\in\separpar{2} \iff \rho\in\separpar{2}$.
\eit
Moreover, it was shown in~\cite{schuch04ssr2} that 
the entanglement cost of $\separpar{2}$ converges to
that of $\separpar{2'}$, so that asymptotically both 
definitions are equivalent.

% S1 no, [S1] tampoco
On the other hand, $\separpar{1}$ and $\equivsep{1}$
do not show the same stability,
although the corresponding individual separability is 
a necessary condition for the separability of the multiple copies
state.

% Pero [S1] => una copia tb S!
\bit
\item
$\rho^{\otimes 2}\in\separpar{1} \Rightarrow \rho\in\separpar{1}$,
\item
$\rho^{\otimes 2}\in\equivsep{1} \Rightarrow \rho\in\equivsep{1}$.
\eit

It is also possible to prove (see section~\ref{subsec:mult}) 
that
\bit
\item
$\rho^{\otimes 2}\in\equivsep{1} \Rightarrow \rho\  \mathrm{PPT}$.
\eit

Therefore, an NPPT state $\rho$ 
is also non
separable according to the broadest definition $\equivsep{1}$
when one takes several copies.
This is true, in particular, for distillable 
states~\cite{bennet96distil,horodecki98distil}.
This suggests that the differences between the various definitions of 
separability may vanish in the asymptotic regime.
The strict equivalence of the classes in this limit, however, is proved
only for the case of $1\times 1$ modes, as detailed in the following section.

%% \subsection{$1\times1$-modes system}
%% % Caso 1x1 => [S1]^2 <=> S2'

%% For the case of a $1\times1$-modes system it is possible to show that
%% $$
%% \rho^{\otimes 2}\in\equivsep{1} \iff \rho\in\separpar{2'}.
%% $$
%% Therefore, in this case all the definitions of entanglement
%% converge when we look at a large number of copies.
%\section{Multiple copies}
%\label{multiple}
% Stable criteria,
% OPEN: S1tilde???

\section{1$\times$1 modes}
% Illustrating it in the 1x1 case
% The complete characterization of all sets in this case
\label{sec:1x1}
%%%%%%%%%%%%%%%%%%%%%%%%%%%%%%%%%%%%%%%%%%%%%%%%%%%%%%%%%%%%%%%%
% Table: Complete characterizations for 1x1 modes
\begin{table*}
\begin{tabular}{|c|c|}
\hline
General &
$\rho=\left (
\begin{array}{c c c c}
1-x-y+z & p & q & r \\
p^* & x-z & s & t \\
q^* & s^* & y-z & w \\
r^* & t^* & w^* &z 
\end{array}
\right )$
\\
\hline
\product{1} &
$z=x\, y$
\\
\hline
\product{2} &
$
\begin{array}{c c c}
\rho=\rho_A\otimes\rho_B &\mathrm {i.e.}&
\left\{
\begin{array}{c}
z=x\, y\\
y\, p=(1-y)w\\
x\, q=(1-x) t\\
x\,y\,r=t\,w\\
x\,y\,s=t\,w^*
\end{array} \right.
\end{array}
$
\\
\hline
\product{3} &
$
\begin{array}{c}
r=s=0\\
z=x\, y\\
\begin{array}{c c}
\left\{
\begin{array}{c}
q=t=0 \\
(1-y)w=-y\, p
\end{array}
\right\}
\, \mathrm{or}\, 
\left\{
\begin{array}{c}
p=w=0 \\
(1-x) t=x\, q
\end{array}
\right\}
\end{array}
\end{array}
$
\\
\hline
$\pset$
&
$p=q=t=w=0$\\
\hline
$\begin{array}{c}
\separpar{1}\\
\separpar{2'}
\end{array}$
&
$\begin{array}{c}
|r|^2\leq (x-z)(y-z)\\
|s|^2\leq z (1-x-y+z)
\end{array}$
\\
\hline
\separpar{2}=\separpar{3}&
$r=s=0$
\\
\hline
\equivsep{1}&
All $\rho \geq 0$
\\
\hline
\end{tabular}
\caption{Characterization of the sets for a $1\times 1$-modes system.}
\label{tab:1x1}
\end{table*}
%%%%%%% End of the TABLE

% The complete characterization

In the case of a small system of only two modes, it is possible 
to apply all the definitions above to the most general density matrix 
and find the complete characterization of each of the sets.
Table~\ref{tab:1x1} shows this characterization.

A generic state of a 1$\times$1-mode system can be written in the
Fock representation as
\beq
\rho=\left (
\begin{array}{c c c c}
1-x-y+z & p & q & r \\
p^* & x-z & s & t \\
q^* & s^* & y-z & w \\
r^* & t^* & w^* &z 
\end{array}
\right ),
\label{gen1x1}
\eeq
where $x,\, y,\, z$ are real parameters, and
with the additional restrictions that ensure $\rho\geq 0$, 
which include $z\leq x,\,y$, and $1+z \geq x+y$.

States in $\product{1}$ must satisfy a single relation between 
expectation values, namely
$\langle c_1  c_2 c_3 c_4\rangle = \langle  c_1 c_2\rangle \langle c_3 c_4 \rangle$,
which reads, in terms of the given parametrization,
$z=x\, y$.
This condition is also necessary for states in $\product{2}$ 
or $\product{3}$.

If a state is in $\product{2}$, it can be written as the tensor product of two 
1-mode matrices, each of them determined by one real and one complex parameter.
This imposes a number of restrictions on the general parameters above, that 
can be read in Table~\ref{tab:1x1}.
Since $\separ{2}$ corresponds to separability in the
isomorphic qubit system, a state will be in $\separ{2}$ iff it has
PPT~\cite{horodecki96}.

%%% P3!!!!
According to~\cite{moriya05}, a state in $\product{3}$
has zero expectation value for all observable products 
$A_{\antipi} B_{\antipi}$, and one of the restrictions of 
$\rho$ to the subsystems is odd with respect to the parity 
transformation.
There are then two generic forms of a product state $\product{3}$ 
in this system, as shown in the table.

If we restrict the study to physical states, i.e. 
those commuting with $\parity$,
the density matrix has a block diagonal structure,
and the most general even $1\times1$ state can be written
\beq
\rho=\left(
\begin{array}{c c c c}
1-x-y+z & 0 & 0 & r\\
0 & x-z & s & 0 \\
0 & s^* & y-z & 0 \\
r^* & 0 & 0 & z
\end{array}
\right).
\label{gen1x1even}
\eeq
Particularizing the conditions for general product states to this form of
the density matrix, where $p=q=t=w=0$, gives the explicit 
characterization of the physical 
product states according to each definition.

In particular, the state~(\ref{gen1x1even}) is in $\productpar{1}$ iff 
$z=x\, y$.
Convex combinations of this kind of states will produce 
density matrices that fulfill $|s|^2\leq z (1-x-y+z)$ 
and $|r|^2\leq(x-z)(y-z)$, and thus have PPT.
This shows that, for this small system,
$\separpar{1}\equiv \separpar{2'}$.

The independent separability of both blocks of $\rho$, that determines 
separability according to $\separpar{2}$, requires that $r=s=0$, i.e. that 
the density matrix is diagonal in this basis.

Finally, the characterization~(\ref{charactZ1})
of $\equivsep{1}$ applied to~(\ref{gen1x1}) yields the
condition that the diagonal of $\rho$ is separable according 
to the tensor product,
so that all even states of $1\times1$-modes are 
in $\equivsep{1}$.

% Multiple copies Case 1x1 => [S1]^2 <=> S2'
If we look at several copies of such
a $1\times1$-modes system,
it is possible to show that
$$
\rho^{\otimes 2}\in\equivsep{1} \iff \rho\in\separpar{2'}.
$$
Therefore, in this case all the definitions of entanglement
converge when we look at a large number of copies.

% show differences in EoF, regions of parameters of particular model.
\subsection{Thermal states of fermionic chains}
% A particular example

All the concepts above can be applied to a particular example.
% With some plots...
%Here we illustrate some of the above concepts with a particular example.
We consider a 1D chain of N fermions subject to the Hamiltonian
\bea
\nn
H&=&\frac{1}{2}\sum_n\left(a_n^{\dagger}a_{n+1}+\mathrm{h.c.}\right) 
-\lambda \sum_n a_n^{\dagger}a_{n} \\
\nn
&&
+\gamma \sum_n\left(a_n^{\dagger}a_{n+1}^{\dagger}+\mathrm{h.c.}\right). 
\eea
This Hamiltonian can be obtained as
the Jordan-Wigner transformation of an XY spin chain with 
transverse magnetic field~\cite{lieb61XY,barouch70XY}.
The Hamiltonian can be exactly diagonalized by means of Fourier 
and Bogoliubov transformations, yielding
$$
H=\sum_{k=-\frac{N-1}{2}}^\frac{N-1}{2}\Lambda_k b_k^{\dagger}b_k,
$$
with $\Lambda_k=\sqrt{\left[\cos\frac{2 \pi k}{N}-\lambda\right]^2+4 \gamma^2 \sin^2\frac{2 \pi k}{N}}$, $b_k=\cos \theta_k a_k+i \sin\theta_k a_{-k}^{\dagger}$,
$\cos 2 \theta_k=\frac{\cos\frac{2 \pi k}{N}-\lambda}{\Lambda_k}$
and $a_k=\frac{1}{\sqrt{N}}\sum_n e^{-i\frac{2 \pi k n}{N}}a_n$.

We consider the thermal state
$\rho=\frac{e^{-\beta H}}{\mathrm{tr}\left[e^{-\beta H}\right]}$,
with inverse temperature $\beta$, and  
calculate the reduced density matrix for two adjacent modes in 
the limit of an infinite chain, by numerical integration of the relevant
expectation values as a function of the three parameters of this model, 
$\lambda$, $\gamma$ and $\beta$.

\begin{figure*}[htp!]
\begin{tabular}{c c}
\includegraphics[width=\columnwidth]{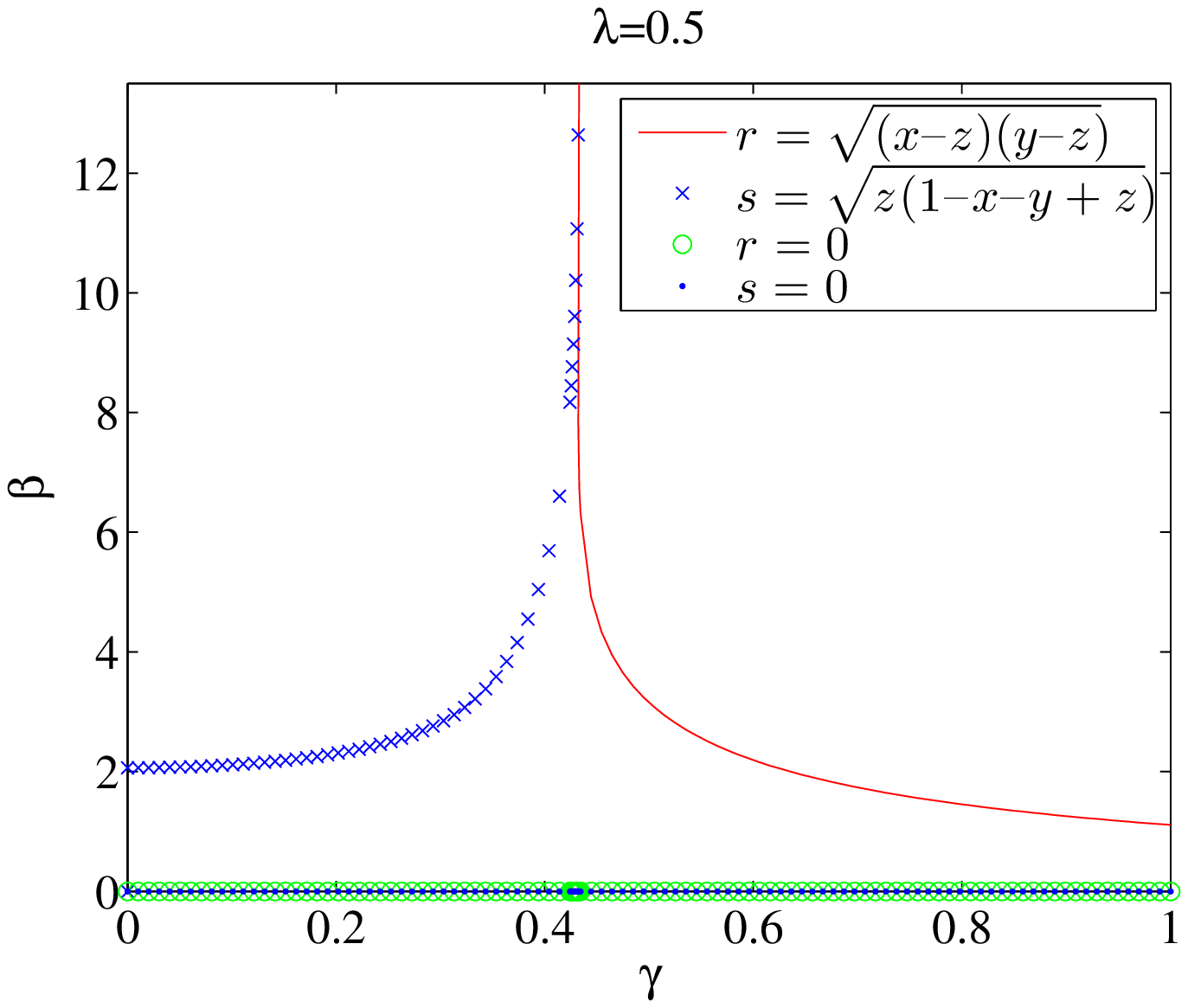}
\includegraphics[width=\columnwidth]{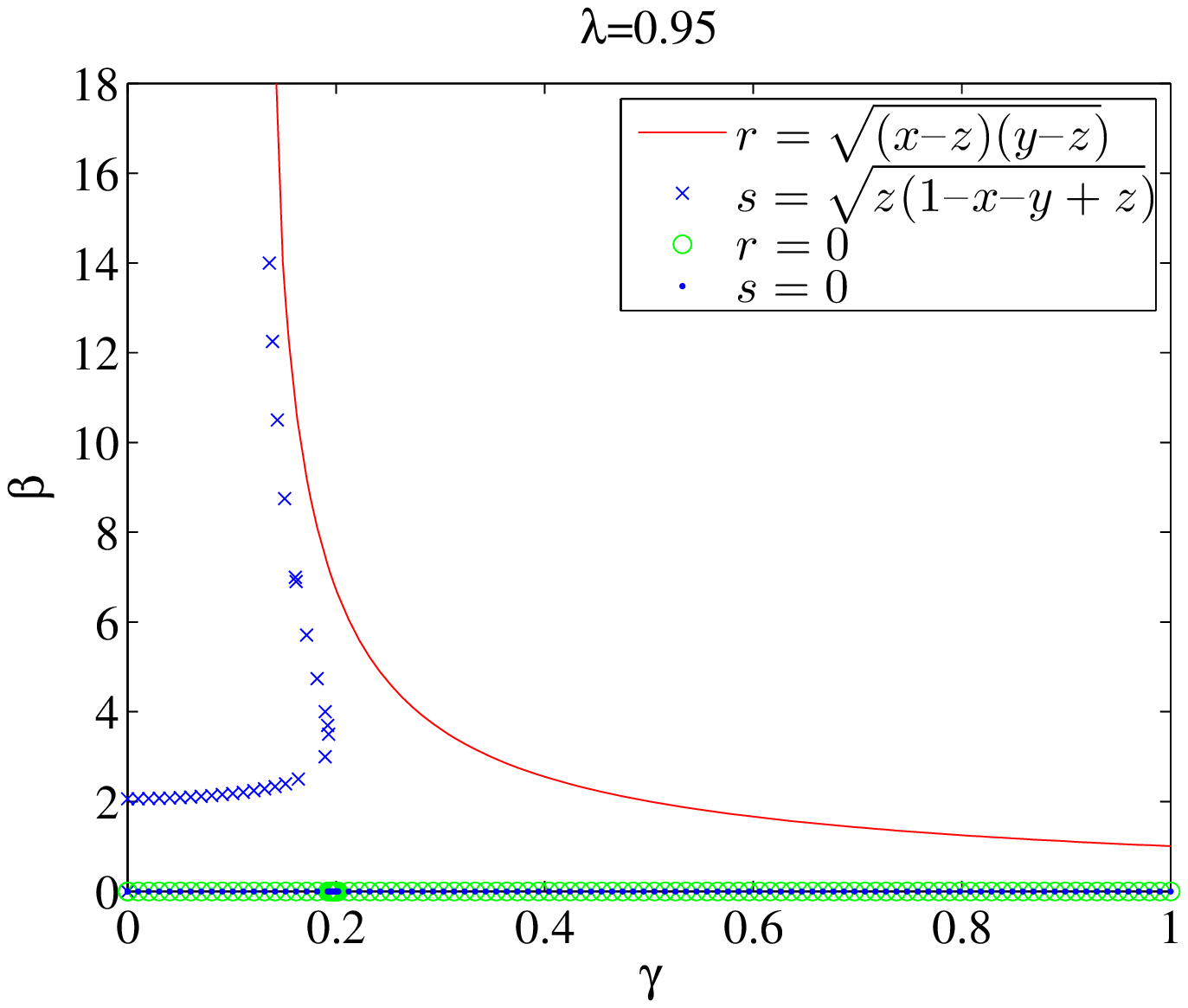}
\end{tabular}
\caption{
Regions of parameters that correspond to separable 
reduced density matrices of two neighboring fermions
according to the different criteria. 
The different curves correspond to values of the parameters for which one
of the conditions of separability (see Table~\ref{tab:1x1}) is satisfied 
with equality.
For a fixed value of $\lambda$ ($\lambda=0.5$ for the left plot, 
$\lambda=0.95$ for the right one), the area at the bottom
corresponds to values of $\beta$,$\gamma$ for which the reduced 
density matrix is in $\separpar{2'}$.
In both cases, the region of $\separpar{2}$ lies on the horizontal axis.
Notice that for $\lambda=0.95$ there is a small range of values of 
$\gamma\lesssim 0.2$
for which the entanglement shows up when increasing the
temperature, as illustrated quantitatively by the figure below.
}
\label{fig:regions}
\end{figure*}

First we may study which values of the parameters make the two modes
entangled according to each of the definitions.
As mentioned above, for a 2-mode system there is no distinction between 
the sets $\separpar{1}$ and $\separpar{2'}$. 
Therefore we look for the limits of the separability regions $\separpar{2'}$
and $\separpar{2}$ for a fixed value of the parameter $\lambda$.
The results are shown in Fig.~\ref{fig:regions}.
For every value of $\lambda$ we may see that the reduced density matrix is
in $\separpar{2}$ only if $\beta=0$, i.e. for all finite values of the 
temperature two adjacent fermions will be entangled according to this 
criterion. 
The region $\separpar{2'}$, on the contrary, changes with the parameters, 
as shown by the plots.

\begin{figure*}[htp!]
\begin{tabular}{c c}
\includegraphics[width=\columnwidth]{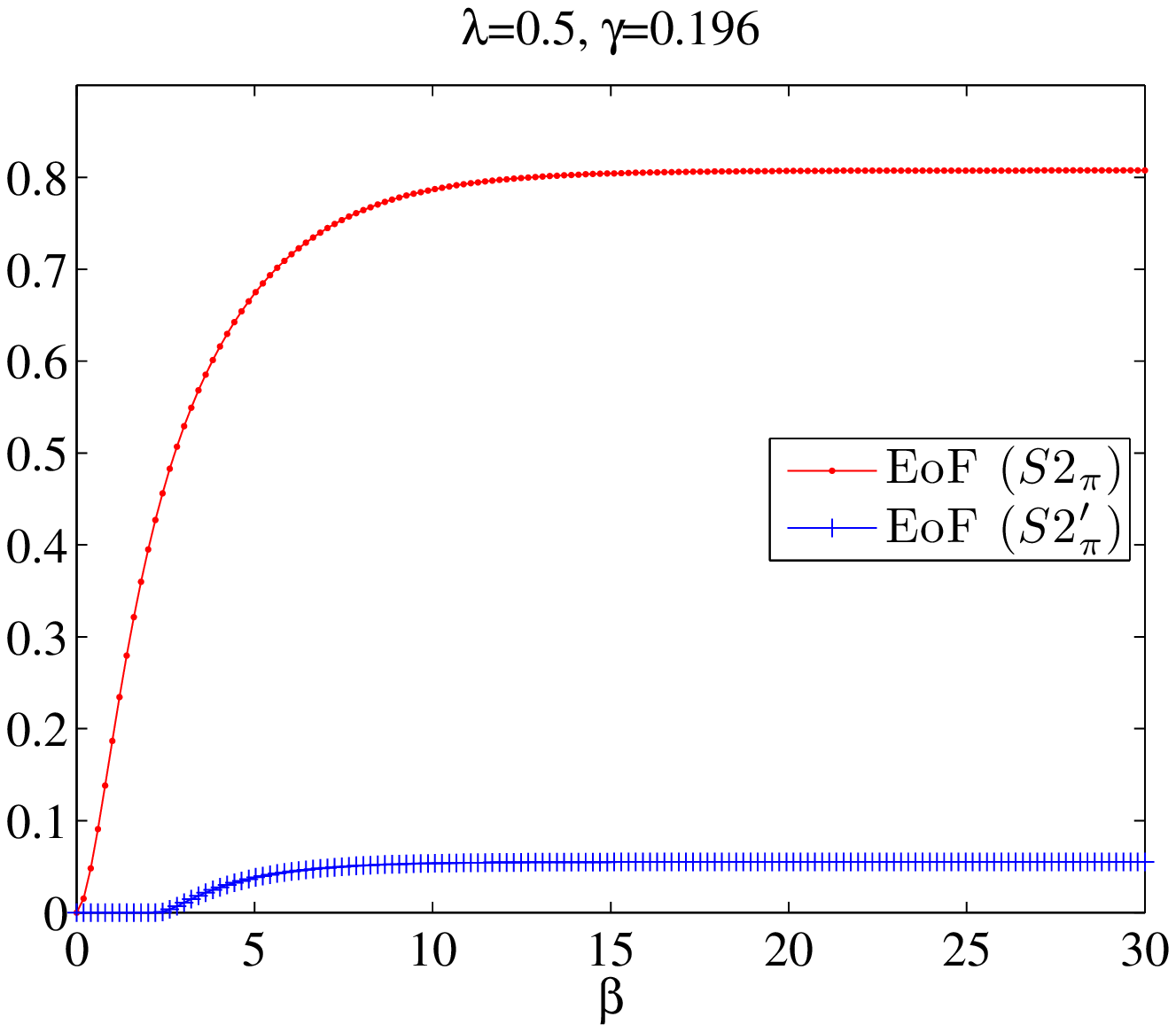}
\includegraphics[width=\columnwidth]{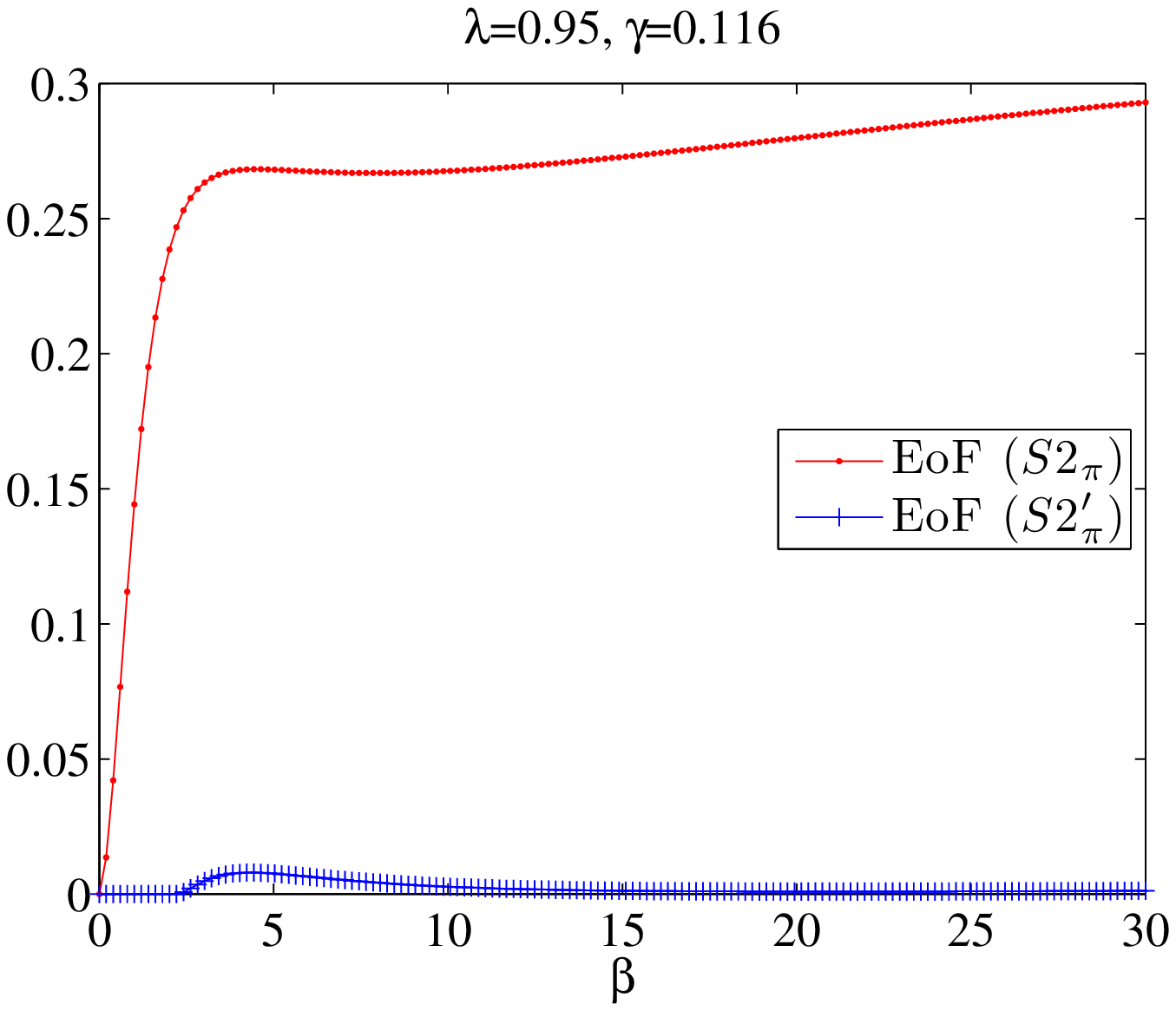}
\end{tabular}
\caption{
Entanglement of formation of the reduced density matrix of 
two neighboring fermions with respect to the sets $\separpar{2}$ 
(dots)
and $\separpar{2'}$ (crosses), for fixed values of $\lambda$ and $\gamma$,
as a function of the inverse temperature.
}
\label{fig:eof}
\end{figure*}

From a quantitative point of view, 
the entanglement with respect to $\separpar{2'}$ can be measured
by the entanglement of formation~\cite{wooters98},
$$
E_F(\rho)=\min_{\{i,\,\psi_i\}} \sum_i p_i E(\psi_i).
$$
With respect to $\separpar{2}$, it is natural to define the 
entanglement of formation conforming to parity conservation
as
$$
E_F^\pi(\rho)=\min_{\{i,\,\psi_i\}} \sum_i p_i E(\psi_i),
$$
where the minimization is performed over ensembles all whose
$\psi_i$ have 
well-defined parity~\cite{schuch04ssr}.
Both quantities can be calculated.
The results as a function 
of the temperature $\beta$, for fixed values of $\lambda$ and $\gamma$, 
are shown in Fig.~\ref{fig:eof}.
Consistently with the results in Fig.~\ref{fig:regions}, there is always
non--zero entanglement with respect to $\separpar{2}$, for $\beta\neq0$.
The entanglement of formation with respect to $\separpar{2'}$
is, for any other value of the temperature, strictly smaller, and 
in fact the reduced density matrix starts to be entangled
at a finite value of $\beta$.

% Reescribir eso.

% The detailed proofs for all the inclusions mentioned
\section{Detailed proofs}
\label{sec:proofs}
%%%%%%%%%%%%%%%%%%%%%%%%%%%%%%%%%%%%%%%%%%%%%%%%%%%%%%%%%%%%%%%%
% Table: Summary of all classes
\begin{table*}
\begin{tabular}{|c|c|c|c|c|}
\hline
\multicolumn{2}{|c|}{Product states} &
\multicolumn{2}{|c|}{Separable states}&
\multirow{2}{*}{Equivalence classes}\\
\cline{1-4}
General
& 
%\multicolumn{1}{p{1.5cm}|}{\centering Physical ($\product{X}\cap\pset$)}
Physical
 &
%\multicolumn{2}{c|}{}
$\mathrm{co}(X)$& 
%\multicolumn{1}{p{1.5cm}|}{\centering Physical ($X\cap\pset$)}
Physical
&
\\
\hline
 \multicolumn{1}{|c}{\multirow{2}{*}{\product{0}}} &
\multicolumn{1}{c|}{}&
$\separ{0}\defeq\mathrm{co}(\product{0})$&
\multirow{2}{*}{$\separpar{0}=\separ{0}\cap\pset$}
&\multirow{2}{*}{$\equivsep{0}$}
\\
\cline{2-3}
&$\productpar{0}\defeq\product{0}\cap\pset$ &
$\separpar{0}\defeq\mathrm{co}(\productpar{0})$&
& % for equiv class
\\

\hline
 \multicolumn{1}{|c}{\multirow{2}{*}{\product{1}}} &
\multicolumn{1}{c|}{}&
$\separ{1}\defeq\mathrm{co}(\product{1})$&
\multirow{2}{*}{$\separpar{1}=\separ{1}\cap\pset$}
&\multirow{2}{*}{$\equivsep{1}$}
\\
\cline{2-3}
&$\productpar{1}\defeq\product{1}\cap\pset$&
$\separpar{1}\defeq\mathrm{co}(\productpar{1})$&
& % for equiv class
\\

\hline
 \multicolumn{1}{|c}{\multirow{2}{*}{\product{2}}} &
\multicolumn{1}{c|}{}&
$\separ{2}\defeq\mathrm{co}(\product{2})$&
$\separpar{2'}\defeq\separ{2}\cap\pset$
&$\equivsep{2'}$
\\
\cline{2-5}
&$\productpar{2}\defeq\product{2}\cap\pset$&
$\separpar{2}\defeq\mathrm{co}(\productpar{2})$&
\separpar{2}
& $\equivsep{2}$ % for equiv class
\\

\hline
 \multicolumn{1}{|c}{\multirow{2}{*}{\product{3}}} &\multicolumn{1}{c|}{}&
$\separ{3}\defeq\mathrm{co}(\product{3})$&
\multirow{2}{*}{$\separpar{3}=\separ{3}\cap\pset$}
&\multirow{2}{*}{$\equivsep{3}$}
\\
\cline{2-3}
&$\productpar{3}\defeq\product{3}\cap\pset$&
$\separpar{3}\defeq\mathrm{co}(\productpar{3})$&
& % for equiv class
\\

\hline
\multicolumn{5}{|c|}{Relations between sets}\\
\hline
%\multicolumn{1}{|p{2cm}|}{\centering $\product{0}=\product{1}$\\
%$\product{2},\product{3}\subset\product{1}$ \\
%$\product{2}\neq \product{3}$
%}&
$
\begin{array}{c}
\product{0}=\product{1}\\
\product{2},\product{3}\subset\product{1}\\
\product{2}\neq \product{3}
\end{array}$
&
%\multicolumn{1}{|p{2cm}|}{\centering $\productpar{0}=\productpar{1}$\\
%$\productpar{2}=\productpar{3}\subset\productpar{1}$}&
$
\begin{array}{c}
\productpar{0}=\productpar{1}\\
\productpar{2}\subset\productpar{1}\\
\productpar{2}=\productpar{3}
\end{array}$
&
%\multicolumn{1}{|p{2cm}|}{\centering $\separ{0}=\separ{1}$\\
%$\separ{3} \subset \separ{2}\subset\separ{1}$}&
$
\begin{array}{c}
\separ{0}=\separ{1}\\
\separ{2}\subset\separ{1}\\
\separ{3} \subset \separ{2}
\end{array}$
&
%\multicolumn{1}{|p{2cm}|}{\centering $\separpar{0}=\separpar{1}$\\
%$\separpar{3}=\separpar{2}\subset\separpar{2'}\subset\separpar{1}$}\\
$
\begin{array}{c}
\separpar{0}=\separpar{1}\\
\separpar{2'}\subset\separpar{1}\\
\separpar{2} \subset \separpar{2'}\\
\separpar{3}=\separpar{2}
\end{array}$
&
$\begin{array}{c}
\equivsep{0}=\equivsep{1}\\
\equivsep{1}=\equivsep{2'}\\
\equivsep{2'}=\equivsep{2}\\
\equivsep{2}=\equivsep{3}
\end{array}$
\\
\hline
\end{tabular}
\caption{Summary of the different definitions of product and separable states.}
\label{tab:summ}
\end{table*}

 % Q: remove this table?
% The detailed proofs for all the inclusions mentioned

This section contains the detailed proofs of all the inclusions 
and equivalences that appear in the text.
Table~\ref{tab:summ} summarizes all the definitions and 
the relations among sets.

\subsection{Product states}
\label{subsec:prod}

\newtheorem{incl}{}[subsection]

\begin{incl}
{$\product{0}\equiv\product{1}$}
\label{proof1}
\end{incl}
\begin{proof}
States in $\product{0}$ satisfy the restriction
that
$$
\rho(A_\pi \, B_\pi) =\tilde{\rho}_A(A_\pi)\tilde{\rho}_B(B_\pi)
$$
for some product state $\tilde{\rho}$ and all parity conserving operators
$A_\pi, \, B_\pi$.
Since the only elements or $\rho$ contributing to such expectation values 
are in the diagonal blocks $\proj{\alpha}{A}\otimes \proj{\beta}{B}
\rho
 \proj{\alpha}{A}\otimes \proj{\beta}{B}$, ($\alpha,\,\beta=e,\,o$), 
the condition is equivalent to saying that the sum of these blocks is equal to 
the (parity commuting) product state 
$\tilde{\rho}=\tilde{\rho}_A\otimes\tilde{\rho}_B$.

The condition for $\rho\in\product{1}$ turns out to be equivalent to this.
We may decompose the state as a sum
$$
\rho=\sum_{\alpha,\,\beta=e,o}\proj{\alpha}{A}\otimes \proj{\beta}{B}
\rho
 \proj{\alpha}{A}\otimes \proj{\beta}{B}+R \defeq \rho'+R,
$$
where $\rho'$ is a density matrix commuting 
with $\parity_A$ and $\parity_B$, and
$R$ contains only the terms that violate parity in some subspace.
%Since any parity commuting operator can be written as
%$A_\pi=\proj{e}{A}A_\pi\proj{e}{A}+\proj{o}{A}A_\pi\proj{o}{A}$, 
It is easy to check that $R$ gives no contribution to expectation 
values of the form $\rho(A_\pi \, B_\pi)$,
so that $\rho'(A_\pi \, B_\pi)=\rho'(A_\pi)\rho'(B_\pi)$.
On the other hand, an operator that is odd under parity has the form
$A_{\antipi}=\proj{e}{A}A_{\antipi}\proj{o}{A}+\proj{o}{A}A_{\antipi}\proj{e}{A}$.
Therefore
$\rho'(A_{\antipi} \, B_{\antipi})=0=\rho'(A_{\antipi})\rho'(B_{\antipi})$.
Since $\rho'$ commutes with parity, and then odd observables give zero
expectation value,
we have checked that all expectation values $\rho'(A\,B)$ factorize 
and then $\rho'$ is a product. 
%Nota: esto en realidad es porque P3pi=P2pi. Comentar?
\end{proof}

\begin{incl}
{$\product{2}\subset\product{1}$}
\label{proof2}
\end{incl}
\begin{proof}
The inclusion $\product{2}\subseteq\product{1}$ is immediate 
from the fact that the 
products of even observables in the 
$A_\pi\, B_\pi$
correspond, via a Jordan-Wigner transformation, to 
products of local even operators $\tilde{A}_e\,\tilde{B}_e$
in the Fock representation, and thus they factorize for any state in
$\product{2}$.
The strict character of the inclusion is shown with an explicit example as
$\PinPiisym$, in Table~\ref{tab:prod}.
\end{proof}

\begin{incl}
{$\product{3}\subset\product{1}$}
\end{incl}
\begin{proof}
The  inclusion $\product{3}\subseteq\product{1}$ in immediate 
from the definitions of both sets.
The example $\PinPiisym\notin\product{3}$ 
(Table~\ref{tab:prod}) shows it is strict.
\end{proof}

\begin{incl}
{$\product{2}\neq\product{3}$}
\end{incl}
\begin{proof}
The example 
\bea
\nn
\PiinPiiisym &=&\PiinPiii \\
\nn
&=&
\frac{1}{2}\left (
\begin{array}{r r}
1 & -1 \\
-1 & 1
\end{array}
\right ) \otimes 
\frac{1}{2}\left (
\begin{array}{r r}
1 & 1 \\
1 & 1
\end{array}
\right ),
\eea
fulfills 
$\PiinPiiisym \in \product{2}$, but $\PiinPiiisym \notin \product{3}$
because it has non vanishing expectation value for products of odd 
operators, f.i.
$\langle c_2 c_3 \rangle_{\PiinPiiisym}=i\neq 0$.

On the other hand, it is also possible to construct a state as
$$
\PiiinPiisym =\PiiinPii,
$$
satisfying
$\PiiinPiisym \in \product{3}$ (it is easy to check the explicit 
characterization for $1\times1$ modes of Table~\ref{tab:1x1}), 
but $\PiiinPiisym \notin \product{2}$ because it is not possible 
to write it as a tensor product.
\end{proof}

\begin{incl}
{$\productpar{2}\subset\productpar{1}$}
\end{incl}
\begin{proof}
The non strict inclusion is immediate from the result for general 
states~(\ref{proof2}).
Actually, the same example $\PinPiisym$ is parity preserving and then 
shows the non equivalence of both sets.
\end{proof}

\begin{incl}
{$\productpar{2}\equiv\productpar{3}$}
\end{incl}
\begin{proof}
For any physical state $[\rho,\,\parity]=0$, the expectation value of
any odd operator is null.
On the other hand, all $\product{3}$ states (in particular those in 
$\productpar{3}$) fulfill 
$\rho(A_{\antipi} B_{\antipi})=0$~\cite{moriya05}.
Since a state in $\productpar{2}$ can be written as a product of two factors
each of them commuting with the local parity operator,
then the only non--vanishing expectation values in these 
sets of states correspond to 
products of parity conserving local observables.
It is then enough to check that
$$
\rho(A_{\pi} B_{\pi})=\rho(A_{\pi})\rho(B_{\pi})
\iff
\rho=\rho_A \otimes \rho_B.
$$
Given the state $\rho$ we can look at the Fock 
representation and write it as an expansion in the Pauli operator basis,
where coefficients correspond to expectation values of products 
$\sigma_{a_1}^{(1)}\otimes \ldots \otimes \sigma_{a_m}^{(m)}$.

Making use of the Jordan-Wigner transformation~(\ref{eq:JW}), 
any product 
of even observables in the Fock space is mapped to a product of even
operators in the subalgebras $\cal{A}$, $\cal{B}$.
So it is easy to see that the property of factorization is equivalent
in both languages and thus
$$
\rho\in\productpar{2}\iff\rho\in\productpar{3}.
$$

This equivalence implies also that of the convex hulls,
$\separpar{2}\equiv\separpar{3}$.
\end{proof}

\subsubsection{Pure states}

\begin{incl}
{For pure states $\productpar{1}\iff\productpar{2}$}
\end{incl}
\begin{proof}

A pure state $|\Psi\rangle\langle\Psi| \in \pset$ is 
such that $\parity \Psi= \pm\Psi$.
We consider the even case (the same reasoning applies for the odd one). 
Since such a state vector is a direct sum of two components,
one of them even with respect to both $\parity_A,\,\parity_B$ 
and the other one odd 
with respect to both local operations,
and applying the Schmidt decomposition to each of those components,
it is always possible to write the state as
$$
|\Psi \rangle =\sum_i \alpha_i |e_i\rangle |\varepsilon_i\rangle +
\sum_i \beta_i|o_i\rangle |\theta_i\rangle,
$$
where $\{|e_i\rangle\}$ ($\{|\varepsilon_i\rangle\}$) are mutually 
orthogonal states with $\parity_A |e_i\rangle=+|e_i\rangle$ 
($\parity_B |\varepsilon_i\rangle=+|\varepsilon_i\rangle$)
and $\{|o_i\rangle\}$ ($\{|\theta_i\rangle\}$)
are mutually 
orthogonal states with $\parity_A |o_i\rangle=-|o_i\rangle$ 
($\parity_B |\theta_i\rangle=-|\theta_i\rangle$).

The condition of $\productpar{1}$ imposes that 
$\langle \Psi |A_\pi\,B_\pi|\Psi\rangle=
\langle \Psi |A_\pi|\Psi\rangle \langle \Psi |B_\pi|\Psi\rangle$ for all
parity preserving observables.
In particular, we may consider those of the form
\bea
A_\pi&=&\sum_k A_k^e|e_k\rangle\langle e_k|+A_k^o |o_k\rangle\langle o_k|,
\nonumber
\\
B_\pi&=&\sum_k B_k^e|\varepsilon_k\rangle\langle \varepsilon_k|+B_k^o |\theta_k\rangle\langle \theta_k|.
\nonumber
\eea
On these observables the restriction reads
\bea
\left(
\sum_i |\alpha_i|^2 A_i^e 
\right.\!\!
&+& \!\!\left.
\sum_i |\beta_i|^2 A_i^o
\right)
\!\left(
\sum_i |\alpha_i|^2 B_i^e\! +\! \sum_i |\beta_i|^2 B_i^o
\right)
\nonumber
\\
&&=\sum_i |\alpha_i|^2 A_i^e B_i^e + \sum_i |\beta_i|^2 A_i^o B_i^o
\nonumber
\eea
Let us assume that the state $\Psi$ has more than one term in the 
even-even sector, i.e. $\alpha_1\neq 0$ and $\alpha_2\neq 0$ (we may
reorder the sum, if necessary). 
Then we apply the condition to $A=A_1^e|e_1\rangle\langle e_1|$,
$B=B_2^e|\varepsilon_2\rangle\langle \varepsilon_2|$, and applying the equality we deduce
$|\alpha_1|^2 A_1^e |\alpha_2|^2 B_2^e=0$, and thus $|\alpha_1| |\alpha_2|=0$,
so that there can only be a single term in the 
$|e_i\rangle |\varepsilon_i\rangle$ sum.
An analogous argument shows that also the sum of 
$|o_i\rangle |\theta_i\rangle$ must have at most one single
contribution, for the state to be in $\productpar{1}$.

By applying the equality to operators
$A=A_1^o|o_1\rangle\langle o_1|$  and
$B=B_1^e|\varepsilon_1\rangle\langle \varepsilon_1|$
we also rule out the possibility that $\Psi$ has a contribution from 
each sector.
Then, if $\Psi\in\productpar{1}$, it has one single term in the Schmidt 
decomposition, and therefore it is a product in the sense of $\productpar{2}$.
\end{proof}

\subsection{Separable states}
\label{subsec:separ}

\begin{incl}
{$\separ{2}\subset\separ{1}$ and $\separpar{2'}\subset\separpar{1}$}
\label{proof4}
\end{incl}
\begin{proof}
The first (non strict) inclusion is immediate from the relation between 
product states~\ref{proof2}.
To see that both sets are not equal, we use again an explicit example.
It is possible to construct a state
in $\productpar{1}\subset\separpar{1}$ which has non-positive partial 
transpose and is thus not in $\separ{2}$.
However, this has to be found in bigger systems than the previous
counterexamples, as in a 2-mode system the conditions for $\separpar{1}$ and 
$\separpar{2}$ are identical, as shown in Table~\ref{tab:1x1}.

% Sketch the procedure?
By constructing random matrices $\rho_A \otimes \rho_B$ in the 
parity preserving sector, and adding off-diagonal terms $R$ which are
also randomly chosen, we 
find a counterexample $\SinSiisym$ in a $2\times 2$-system
such that $\SinSiisym \in \productpar{1}$ by construction, but
its partial transposition with respect to the subsystem B, 
$\SinSiisym^{T_B}$, has a negative eigenvalue.

% El estado no cabe
%\SinSii

When taking intersection with the set of physical states, the inclusion 
still holds, and it is again strict, since the counterexample 
$\SinSiisym$ is in particular in $\productpar{1}$.
\end{proof}

%% Include equivalences type $\separpar{1}\equiv \separ{1} \cap \pset$???

\begin{incl}
{$\separpar{1}\equiv\separ{1} \cap \pset$}
\end{incl}
\begin{proof}
Obviously,
$\separpar{1}\subseteq\separ{1} \cap \pset$.
To see the converse direction of the inclusion, we consider a state
$\rho\in\separ{1}\cap\pset$. Then there is a decomposition
$\rho=\sum_i\lambda_i\rho_i$ with $\rho_i\in\product{1}$, but not
necessarily in $\pset$.
We may split the sum into the even and odd terms under the parity 
operator,
$$
\rho=\rho_{\pi}+\rho_{\antipi}\defeq\sum_i \lambda_i \frac{1}{2}(\rho_i+\parity \rho_i \parity)
+\sum_i \lambda_i \frac{1}{2}(\rho_i-\parity \rho_i \parity).
$$
The second term, $\rho_{\antipi}$, gives no contribution to operators that commute 
with $\parity$. Since $\rho$ is physical, this term also gives zero 
contribution to odd observables, so that
$$
\rho=\sum_i \lambda_i \frac{1}{2}(\rho_i+\parity \rho_i \parity).
$$
It only remains to be shown that each 
$\rho_{i_\pi}\defeq \frac{1}{2}(\rho_i+\parity \rho_i \parity)$ is still a product state
in $\productpar{1}$.
But since for parity commuting observables all the contributions come 
from the symmetric part of the density matrix,
$\rho_i(A_\pi B_\pi)=\rho_{i_\pi}(A_\pi B_\pi)$,
and the condition for $\productpar{1}$ holds for $\rho_{i_\pi}$.
Therefore we have found a convex decomposition of $\rho$ in terms of 
product states all of them conforming to the symmetry.

The analogous relation for $\separpar{2}$ was shown in~\cite{moriya05}.
\end{proof}

\begin{incl}
{$\separpar{2}\subset\separpar{2'}$}
\end{incl}
\begin{proof}
Since $\productpar{2}=\product{2}\cap\pset$, taking convex hulls and 
intersecting again with $\pset$ implies that
$\separpar{2}\subseteq\separpar{2'}$.
However, not all separable states can be decomposed as a convex sum of
product states all of them conforming to the parity symmetry.
In particular, the state
$$
\SiipnSiisym=\SiipnSii,
$$
which has PPT and is thus in $\separpar{2'}$, is not in $\separpar{2}$ 
(recall that for the $1\times1$-system, only density matrices  which are
diagonal in the number basis are in $\separpar{2}$).
\end{proof}

\begin{incl}
{$\equivsep{1}\equiv\equivsep{2'}\equiv\equivsep{2}$}
\end{incl}
\begin{proof}
From the relations
$\separpar{2}\subset\separpar{2'}\subset\separpar{1}$ and the definition
of the equivalence classes it is evident that
$\equivsep{2}\subseteq\equivsep{2'}\subseteq\equivsep{1}$.
To show the equivalence of all sets it is enough to prove that any
 state $\rho \in \equivsep{1}$ is also in $\equivsep{2}$, i.e. that 
there exists a state in $\separpar{2}$ equivalent to $\rho$.

For $\rho\in\equivsep{1}$, there is a $\tilde{\rho} \in \separpar{1}$,
i.e. $\tilde{\rho}=\sum\lambda_k\tilde{\rho}_k$ with each
$\tilde{\rho}_k \in \productpar{1}$,
producing identical expectation values for products of even operators 
$A_\pi B_\pi$.
If we define
$$
\rho_k'\defeq \sum_{\alpha,\,\beta=e,\,o}\proj{\alpha}{A}\otimes\proj{\beta}{B}
\tilde{\rho}_k \proj{\alpha}{A}\otimes\proj{\beta}{B},
$$
it is evident that $\rho'\defeq\sum_k\lambda_k \rho_k'$ produces the 
same expectation values as $\tilde{\rho}$ for the 
relevant operators (see proof~\ref{proof1}).
Therefore, $\rho \sim \rho'$.
Moreover, since $\rho_k'(A_{\antipi}B_{\antipi})=0$ for all odd-odd 
products, every $\rho_k'\in\productpar{2}$, and so
$\rho'\in\separpar{2}$.
\end{proof}

\subsection{Multiple copies}
\label{subsec:mult}

\begin{incl}
{$\rho^{\otimes 2}\in\separpar{1} \Rightarrow \rho\in\separpar{1}$}
\label{mult1}
\end{incl}
\begin{proof}
An arbitrary state can be decomposed in two terms,
$\rho=\rho_E+\rho_O$, where
$$
\rho_E\defeq
\sum_{\alpha,\,\beta=e,\,o} \proj{\alpha}{A}\otimes \proj{\beta}{B}
\rho
 \proj{\alpha}{A}\otimes \proj{\beta}{B},
$$
and
$$
\rho_O\defeq
\sum_{
\substack{
{\alpha,\,\beta,\,\gamma,\,\delta=e,\,o}\\
{(\alpha,\,\beta)\neq (\gamma,\,\delta)}}
} \proj{\alpha}{A}\otimes \proj{\beta}{B}
\rho
 \proj{\gamma}{A}\otimes \proj{\delta}{B}.
$$
For any state in $\separpar{1}$, there exists a decomposition
$\rho_E=\sum_i\lambda_i\rho_E^i$,
$\rho_O=\sum_i\lambda_i\rho_O^i$,
such that $\rho_E^i+\rho_O^i\in\productpar{1}$.
Let us consider two copies of a state such that 
$\tilde{\rho}\defeq\rho^{\otimes 2}\in\separpar{1}$.
Then, using the above decomposition of $\tilde{\rho}$, 
and taking the partial trace with respect to the second
system,
we obtain a decomposition of the single copy,
$\rho=\rho_E+\rho_O=\sum_i \lambda_i \mathrm{tr}_2(\tilde{\rho}_E^i)
+\sum_i \lambda_i \mathrm{tr}_2(\tilde{\rho}_O^i)$.
Since $\tilde{\rho}_E^i$ was a tensor product,
$\tilde{\rho}_E^i=\tilde{\rho}_{\tilde{A}}\otimes\tilde{\rho}_{\tilde{B}}$,
with $\tilde{A}\equiv A_1A_2$, $\tilde{B}\equiv B_1B_2$,
so is  $\mathrm{tr}_2(\tilde{\rho}_E^i)$, and therefore
$\rho\in\separpar{1}$.
\end{proof}

\begin{incl}
{$\rho^{\otimes 2}\in\equivsep{1} \Rightarrow \rho\in\equivsep{1}$}
\label{mult2}
\end{incl}
\begin{proof}
Using the same decomposition as above,
$\rho=\rho_E+\rho_O$,
a state $\rho\in\equivsep{1}$ satisfies $\rho_E\in\separpar{2'}$.
If we consider 
$\tilde{\rho}\defeq\rho^{\otimes 2}=\tilde{\rho}_E+\tilde{\rho}_O$,
the condition $\equivsep{1}$ on the state of the two copies reads
$$
\tilde{\rho}_E=\rho_E\otimes\rho_E+\rho_O\otimes\rho_O\in\separpar{2'},
$$
in terms of the components of the single copy state.
Taking the trace with respect to one of the copies, then, 
and using the fact that $\rho_O$ is traceless,
$\rho_E\in\separpar{2'}$,
so that $\rho\in\equivsep{1}$.
\end{proof}

\begin{incl}
{$\rho\  \mathrm{NPPT} \Rightarrow \rho^{\otimes 2}\notin\equivsep{1}$}
\label{mult4}
\end{incl}
\begin{proof}
We may restrict the proof to states such that  $\rho \in \equivsep{1}$.
In other case the implication follows immediately
from  the previous result (\ref{mult2}).
Written in a basis of well-defined local parities,
any density matrix that commutes with the parity operator has a block 
structure (analogous to that of~(\ref{gen1x1even}) for the $1\times 1$ 
case).
%, where each block has now dimension $2^{m_A+m_B-2}\times2^{m_A+m_B-2}$.
\beq
\rho=\left(
\begin{array}{c c c c}
\rho_{ee} & 0 & 0 & C\\
0 & \rho_{eo} & D & 0 \\
0 & D^{\dagger} & \rho_{oe} & 0 \\
C^{\dagger} & 0 & 0 & \rho_{oo}
\end{array}
\right).
\label{geneven}
\eeq
The diagonal blocks correspond to the projections onto simultaneous 
eigenspaces of both parity operators,
$\rho_{\alpha\beta}=\proj{\alpha}{A}\otimes \proj{\beta}{B} \rho \proj{\alpha}{A}\otimes \proj{\beta}{B}$,
whereas
$C=\proj{e}{A}\otimes \proj{e}{B} \rho \proj{o}{A}\otimes \proj{o}{B}$
and $D=\proj{e}{A}\otimes \proj{o}{B} \rho \proj{o}{A}\otimes \proj{e}{B}$.

From the characterization~(\ref{charactZ1}) of separability,
the state is in \equivsep{1} iff
all the diagonal blocks 
$\rho_{\alpha \beta}$ are in \separpar{2'}.
It is then enough to prove that the partial transpose of $\rho$ 
is positive iff 
$\proj{e}{A}\otimes \proj{e}{B}
\rho\otimes\rho
 \proj{e}{A}\otimes \proj{e}{B}$
has PPT.
Non positivity of the partial transpose of $\rho$ implies then 
the non separability (\separpar{2'})
of the one of the diagonal blocks of $\rho\otimes\rho$.

The partial transposition of the above matrix yields
\beq
\rho^{T_B}=\left(
\begin{array}{c c c c}
\rho_{ee}' & 0 & 0 & D'\\
0 & \rho_{eo}' & C' & 0 \\
0 & (C'^{\dagger}) & \rho_{oe}' & 0 \\
(D')^{\dagger} & 0 & 0 & \rho_{oo}'
\end{array}
\right),
\label{genevenPT}
\eeq
where $X'\defeq X^{T_B}$, and the $T_B$ operation acts on each block
transposing the last $m_B-1$ indices.

If we take two copies of the state, we find for the corresponding
uppermost diagonal block
$\tilde{\rho}_{ee}\defeq \proj{e}{A}\otimes \proj{e}{B}
\rho\otimes\rho
 \proj{e}{A}\otimes \proj{e}{B}$,
\beq
\tilde{\rho}_{ee}=\left(
\begin{array}{c c c c}
\rho_{ee}\otimes\rho_{ee} & 0 & 0 & C\otimes C\\
0 & \rho_{eo}\otimes\rho_{eo} & D\otimes D & 0 \\
0 & D^{\dagger}\otimes  D^{\dagger} & \rho_{oe}\otimes\rho_{oe} & 0 \\
C^{\dagger}\otimes C^{\dagger} & 0 & 0 & \rho_{oo}\otimes\rho_{oo}
\end{array}
\right),
\label{rho2ee}
\eeq 
and for the partial transposition
\beq
(\tilde{\rho}_{ee})^{T_B}=\left(
\begin{array}{c c c c}
\rho_{ee}'\otimes\rho_{ee}' & 0 & 0 & D'\otimes D'\\
0 & \rho_{eo}'\otimes\rho_{eo}' & C'\otimes C' & 0 \\
0 & C'^{\dagger}\otimes  C'^{\dagger} & \rho_{oe}'\otimes\rho_{oe}' & 0 \\
D'^{\dagger}\otimes D'^{\dagger} & 0 & 0 & \rho_{oo}'\otimes\rho_{oo}'
\end{array}
\right).
\label{rho2eePT}
\eeq 
The matrices (\ref{genevenPT}) and (\ref{rho2eePT}) are the direct sum of two
blocks. Thus they are positive definite iff each such block is positive 
definite.
Let us consider one of the blocks of (\ref{rho2eePT}), namely
\beq
\left(
\begin{array}{c c}
\rho_{ee}'\otimes\rho_{ee}'& D'\otimes D'\\
D'^{\dagger}\otimes D'^{\dagger} & \rho_{oo}'\otimes\rho_{oo}'
\end{array}
\right).
\label{B1rho2ee}
\eeq 
%Applying Lemma 3.5.12 of~\cite{zhang91}, and 
% Cite Bathia??
Let us first assume that $\rho_{oo}'$ is non-singular.
Applying a standard theorem in matrix analysis 
%(see f.i.~\cite{zhang91}, Thm. 6.13) 
and making use of the fact that our 
$\rho \in \equivsep{1}$, so that each diagonal block is PPT,
we obtain that~(\ref{B1rho2ee}) is positive iff 
$$
\rho_{ee}'\otimes\rho_{ee}'\geq (D'\otimes D') (\rho_{oo}'^{-1}\otimes
\rho_{oo}'^{-1}) (D'^{\dagger}\otimes D'^{\dagger}),
$$
which holds iff
$$
\rho_{ee}' \geq D' (\rho_{oo}')^{-1}D'^{\dagger}.
$$
Reasoning in the same way for the second block of (\ref{rho2eePT}),
one gets that
\beq
(\tilde{\rho}_{ee})^{T_B} \geq 0 \Leftrightarrow \rho^{T_B}\geq 0.
\eeq
The result holds also if the assumption of non-singularity 
of $\rho_{oo}$ ($\rho_{oe}$ for the second block) is not valid.
In that case, we may take $\rho_{oo}$ diagonal and then, 
by positivity of $(\tilde{\rho}_{ee})^{T_B}$ (or $\rho^{T_B}$
for the reverse implication), find that $D'$ must have some 
null columns. This allows us to reduce both matrices to a similar 
block structure, where the reduced  $\rho_{oo}$ ($\rho_{oe}$) is 
non-singular.
\end{proof}

\begin{incl}
{For $1\times1$ systems, $\rho^{\otimes 2}\in\equivsep{1} \iff \rho\in\separpar{2'}$}
\label{mult3}
\end{incl}
\begin{proof}
One of the directions is immediate, and valid for an arbitrarily large system, 
since $\rho\in\separpar{2'}$ implies
$\rho^{\otimes 2}\in\separpar{2'}\subset\separpar{1}\subset\equivsep{1}$.
On the other hand, if we take 
$\tilde{\rho}\defeq\rho^{\otimes 2}\in\equivsep{1}$,
then the diagonal blocks of this state are separable,
 in particular
$\proj{e}{\tilde{A}}\otimes\proj{e}{\tilde{B}}\tilde{\rho}
\proj{e}{\tilde{A}}\otimes\proj{e}{\tilde{B}}\in\separpar{2'}$, 
which was calculated in (\ref{rho2ee}).
For the case of $1\times1$ modes, with 
$\rho$ given by~(\ref{gen1x1even}), this block reads
$$
\left(
\begin{array}{c c c c}
(1-x-y+z)^2 & 0 & 0 & r^2\\
0 & (x-z)^2 & s^2 & 0 \\
0 & (s^*)^2 & (y-z)^2 & 0 \\
(r^*)^2 & 0 & 0 & z^2
\end{array}
\right).
$$
This is in $\separpar{2'}$ iff it has PPT, and 
this happens if and only if $\rho$ has PPT, i.e. $\rho\in\separpar{2'}$.
\end{proof}

\bibliography{fermionsep}

\begin{thebibliography}{32}
\bibitem[{foo({\natexlab{a}})}]{foot1}
\bibinfo{note}{Yet a different definition is based on a generalized notion of
  entanglement which is relative to distinguished subspaces of
  observables~\cite{barnum04}.}
\bibitem[{foo({\natexlab{b}})}]{foot2}
\bibinfo{note}{In the following we use for the expectation values the notation
  $\rho(X)\defeq\mathrm{tr}(\rho\,X)$.}
\end{thebibliography}

\end{document}